\documentclass[aps,preprint,showpacs,groupedaddress,floatfix]{revtex4}
\usepackage{graphicx}
\usepackage{dcolumn}
\usepackage{bm}

\begin{document}
\title{The Proton Electric Pygmy Dipole Resonance}
\author{N. Paar}
\affiliation{Institut f\" ur Kernphysik, Technische Universit\" at Darmstadt, Schlossgartenstrasse 9,
D-64289 Darmstadt, Germany}
\author{D. Vretenar}
\affiliation{Physics Department, Faculty of Science, University of Zagreb, Croatia}
\author{P. Ring}
\affiliation{Physik-Department der Technischen Universit\"at M\"unchen, D-85748 Garching,
Germany}
\date{\today}

\begin{abstract}
The evolution of the low-lying E1 strength in proton-rich 
nuclei is analyzed in the framework of the self-consistent
relativistic Hartree-Bogoliubov (RHB) model and the 
relativistic quasiparticle random-phase approximation (RQRPA).
Model calculations are performed for a series of N=20 isotones 
and Z=18 isotopes. For nuclei close to the proton drip-line, the
occurrence of pronounced dipole peaks is predicted in the 
low-energy region below 10 MeV excitation energy. 
From the analysis of the proton and neutron transition densities 
and the structure of the RQRPA amplitudes, it is shown that 
these states correspond to the proton pygmy dipole resonance.    
\end{abstract}

\maketitle

A number of experimental and theoretical studies of the multipole 
response of exotic nuclei far from stability have been reported in 
recent years. On the neutron-rich side, in particular, the 
possible occurrence of the pygmy dipole resonance (PDR), 
i.e. the resonant oscillation of the weakly-bound neutron 
skin against the isospin saturated proton-neutron core, has 
been investigated. The onset of low-lying E1 strength 
has been observed not only in exotic nuclei with a large neutron 
excess, e.g. for neutron-rich oxygen isotopes \cite{Lei.01}, 
but also in stable nuclei with moderate proton-neutron asymmetry,
like $^{44,48}$Ca and $^{208}$Pb~\cite{Rye.02,End.03,Har.04}.
On the theory side, various
models have been employed in the investigation of the nature of 
the low-lying dipole strength. Recent studies include the application
of the Skyrme Hartree-Fock + quasiparticle  RPA
with phonon coupling~\cite{Col.01,Sarchi},
the time-dependent density-matrix theory~\cite{Toh.01}, the continuum
linear response in the coordinate-space Hartree-Fock Bogoliubov
formalism~\cite{Mat.01}, the quasiparticle phonon
model~\cite{Rye.02,End.03,TLS.04,TLS.04a}, the relativistic 
RPA~\cite{Vrepyg1.01,Vrepyg2.01}, and the relativistic quasiparticle RPA
(RQRPA) \cite{Paar.03,Paar.05}. The PDR is interesting not only as 
a new and exotic mode of nuclear excitations, but it also plays an
important role in predictions of
neutron capture rates in the r-process nucleosynthesis, 
and consequently in the calculated elemental abundance distribution.
Even though the E1 strength of the PDR is small compared to the 
total dipole strength, the occurrence of the PDR significantly
enhances the radiative neutron capture cross section on neutron-rich
nuclei, as shown in recent large-scale QRPA calculations 
of the E1 strength for the whole nuclear chart \cite{Gor.02,Gor.04}.
 
In the analysis based on the relativistic (Q)RPA 
\cite{Vrepyg1.01,Vrepyg2.01,Paar.03,Paar.05}, it has been shown  
that in neutron-rich nuclei the electric dipole 
response is characterized by the fragmentation of the strength 
distribution and its spreading into the low-energy region.
In light nuclei, the low-lying dipole strength is not collective
and originates from non-resonant single-neutron
excitations. In medium-heavy and heavy neutron-rich nuclei, 
on the other hand, some of the low-lying dipole states  
display a more distributed structure of the RQRPA amplitudes. 
The study of the corresponding transition densities and velocity
distributions revealed the dynamics of the neutron pygmy dipole 
resonance.  

In this work we employ the relativistic QRPA in the study 
of the evolution of low-lying dipole strength in 
proton-rich nuclei. Because the proton drip-line is 
much closer to the line of $\beta$-stability than the 
neutron drip-line, bound nuclei with an excess of protons 
over neutrons can be only found in the region of light 
$Z\leq 20$ and medium mass $20 < Z \leq 50$ elements. 
For $Z > 50$, nuclei in the region of the proton drip-line 
are neutron-deficient rather than proton-rich.    
In addition, in contrast to the evolution of the neutron skin 
in neutron-rich systems, because of the presence of the Coulomb 
barrier nuclei close to the proton 
drip-line generally do not exhibit a pronounced proton 
skin, except for very light elements. Since in light nuclei 
the multipole response is generally less collective, all 
these effects seem to preclude the formation of the 
pygmy dipole states in nuclei close to the proton drip-line.
Nevertheless, the present analysis will show that proton 
pygmy dipole states can develop in light and medium mass 
proton-rich nuclei. 

The relativistic QRPA \cite{Paar.03} is formulated in the
canonical single-nucleon basis of the relativistic
Hartree-Bogoliubov (RHB) model and is fully self-consistent.
For the interaction in the particle-hole channel effective Lagrangians with
nonlinear meson self-interactions or density-dependent 
meson-nucleon couplings are used, and pairing correlations are
described by the pairing part of the finite-range Gogny interaction. Both in
the particle-hole and pairing channels, 
the same interactions are used in the RHB
equations that determine the canonical quasiparticle basis, and in the
matrix equations of the RQRPA. This feature is essential for the
decoupling of the zero-energy mode which corresponds to the spurious
center-of-mass motion. In addition to configurations built from
two-quasiparticle states of positive energy, the RQRPA configuration space
contains quasiparticle excitations formed from the 
ground-state configurations of fully or partially occupied
states of positive energy and the empty negative-energy states from the
Dirac sea. In the present analysis of the dipole response of 
proton-rich nuclei in the fully self-consistent RHB+RQRPA framework, 
the density-dependent effective 
interaction DD-ME1~\cite{Nik1.02} is employed in the
$ph$-channel, and the finite range Gogny interaction D1S~\cite{BGG.91}
in the $pp$-channel.

\begin{figure}
\includegraphics[scale=0.6,angle=270]{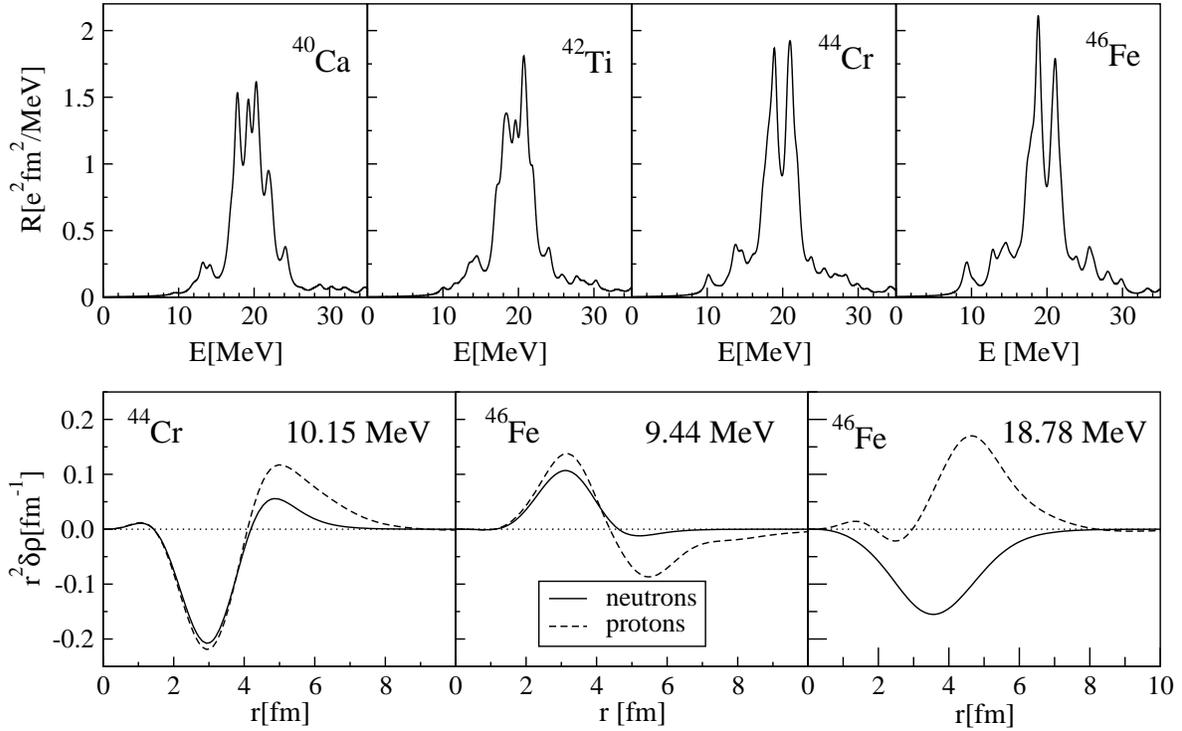}
\caption{The RHB+RQRPA isovector dipole strength distributions 
in the N=20 isotones, calculated with the 
DD-ME1 effective interaction. For $^{44}$Cr and $^{46}$Fe the
proton and neutron transition densities for the main
peak in the low-energy region are displayed in the 
lower panel and, for $^{46}$Fe, the transition densities 
for the main GDR peak.}
\label{fig1}
\end{figure}

In Fig.~\ref{fig1} we display the RQRPA dipole strength 
distributions in the N=20 isotones: $^{40}$Ca, $^{42}$Ti, 
$^{44}$Cr, and $^{46}$Fe. The electric dipole response 
\begin{eqnarray}  
B^T(EJ,\omega_{\upsilon}) & = & \frac{1}{2J_{i}+1}
\bigg\vert \sum_{\mu\mu'} \bigg\{ X^{\upsilon, J0}_{\mu\mu'} \langle
\mu || \hat{Q}^T_J || \mu' \rangle \nonumber \\ 
& + &~(-1)^{j_{\mu}-j_{\mu'}+J} \, Y^{\upsilon, J0}_{\mu\mu'}
\, \langle \mu' || \hat{Q}^T_J || \mu \rangle \,
\bigg\}(u_{\mu}v_{\mu'}+~(-1)^{J}v_{\mu}u_{\mu'})
\bigg\vert ^2 \quad,  
\label{strength}  
\end{eqnarray} 
is calculated for the isovector dipole operator
\begin{equation}
\hat{Q}_{1 \mu}^{T=1} \ = \frac{N}{N+Z}\sum^{Z}_{p=1} r_{p}Y_{1 \mu}
- \frac{Z}{N+Z}\sum^{N}_{n=1} r_{n}Y_{1 \mu} \; .
\end{equation}
For each RQRPA energy $\omega_{\nu}$,
$X^{\nu}$ and $Y^{\nu}$ denote
the corresponding forward- and backward-going two-quasiparticle 
amplitudes, respectively. $v_{\mu}$ and $u_{\mu}$ are the occupation numbers of the
single-particle levels in the canonical basis. The discrete spectra
are averaged with the Lorentzian distribution
\begin{equation}
\label{diplor} R\left( E \right) = \sum_{i}B(E1,1_i \rightarrow 0_f)
\frac{\Gamma/2\pi}{\left(E-E_{i}\right)^2+\Gamma^2/4},
\end{equation}
with $\Gamma = 1$ MeV as an arbitrary choice for the width of the Lorentzian.

The dipole strength distributions in Fig.~\ref{fig1} are dominated by the 
isovector giant dipole resonances (GDR) at $\approx 20$ MeV excitation energy.
In these relatively light systems the GDR still 
exhibits pronounced fragmentation. With the increase of the number of protons,  
low-lying dipole strength appears in the region below the GDR and, 
for $^{44}$Cr and $^{46}$Fe, a pronounced 
low-energy peak is found at $\approx 10$ MeV excitation energy. 
What is the nature of this low-lying dipole state?
In the lower panel of Fig.~\ref{fig1} we plot the proton and neutron 
transition densities for the peaks at 10.15 MeV in $^{44}$Cr and 
9.44 MeV in $^{46}$Fe, and compare them with the transition densities
of the GDR state at 18.78 MeV in $^{46}$Fe. Obviously the dynamics 
of the two low-energy peaks is very different from that
of the isovector GDR: the proton and neutron transition densities
are in phase in the nuclear interior and 
there is almost no contribution from the neutrons
in the surface region. The low-lying state does not belong to
statistical E1 excitations sitting on the tail of the GDR, but
represents a fundamental mode of excitation: the proton electric pygmy 
dipole resonance (PDR).

\begin{figure}
\includegraphics[scale=0.75,angle=0]{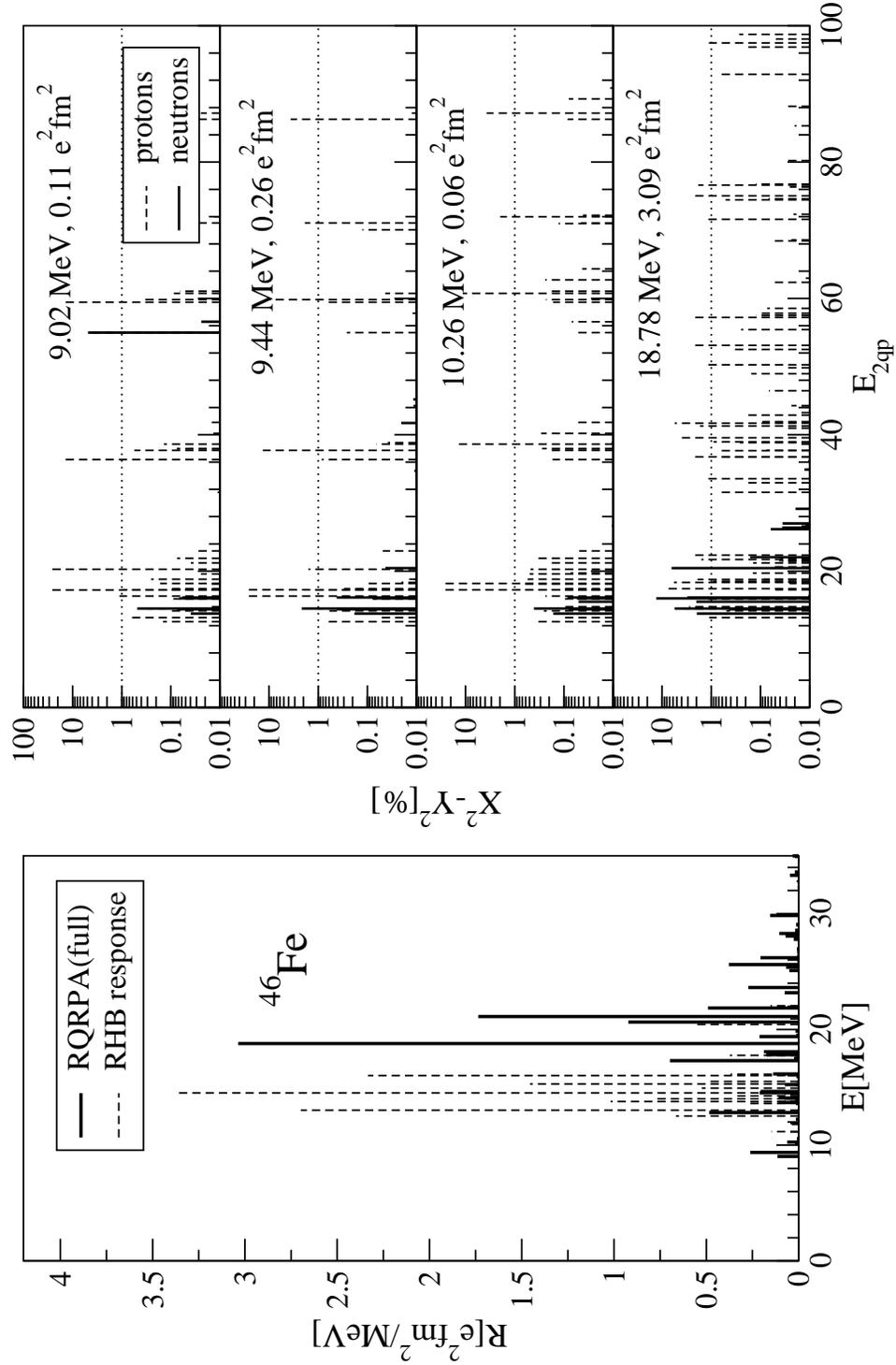}
\caption{The discrete Hartree-Bogoliubov and full RQRPA E1 spectra in 
$^{46}$Fe (left panel). For the main peaks in the low-energy 
region and in the region of the isovector GDR, the distributions of the 
RQRPA amplitudes are shown as functions of the unperturbed 
energy of the respective $2qp$-configurations (right panel).}
\label{fig2}
\end{figure}

In Fig.~\ref{fig2} we analyze the RQRPA structure of the 
dipole response in $^{46}$Fe. This nucleus is located at the 
proton drip-line (see Ref.~\cite{Michael} for a recent review 
on the limits of nuclear stability) and, very recently, for the 
first time evidence for ground-state two-proton radioactivity 
was reported in the decay of $^{45}$Fe \cite{Gio.02,Pfu.02}.
The discrete unperturbed Hartree-Bogoliubov and full RQRPA E1 
spectra of $^{46}$Fe are shown in the left panel. Since the 
interaction is repulsive in the isovector channel, one expects that 
the inclusion of the residual interaction will result in the shift 
of the full RQRPA spectrum to higher energies with respect to the 
unperturbed spectrum. This is clearly the case for the states in 
the GDR region. The lowest group of states at $\approx 10$ MeV, 
however, gains strength and is shifted to lower energy. Obviously 
the nature of these states is not isovector. The QRPA structure of 
these states is shown in the panel on the right, where for the 
three low-lying states at 9.02, 9.44, and 10.26 MeV, as well 
as for the strongest state in the GDR region at 18.78 MeV, 
we plot the corresponding QRPA amplitude of  
proton and neutron {$2qp$} configurations
\begin{equation}
\xi_{2qp}=\left|X^{\nu}_{2qp}\right|^2-
\left|Y^{\nu}_{2qp}\right|^2~,
\end{equation}
with the normalization
condition
\begin{equation}
\sum_{2qp}\xi_{2qp}=1.
\label{norm}
\end{equation}
For each of the four dipole states, in addition to the excitation energy  
we have also included the corresponding B(E1) value. The QRPA
amplitudes are shown in a logarithmic plot as functions of the 
unperturbed energy of the respective $2qp$-configurations.
Only amplitudes which contribute more than 0.01\% are shown, and 
we also differentiate between proton and neutron configurations. 
We note that, rather than a single
proton {$2qp$} excitation, the low-lying states are 
characterized by a superposition of a number of 
{$2qp$} configurations. Obviously the pygmy states display 
a degree of collectivity that can be directly compared with the QRPA 
structure of the GDR state at 18.78 MeV. In addition, proton 
{$2qp$} configurations account for $\approx$ 99\% of the 
QRPA amplitude of the pygmy states, whereas the ratio
of the proton to neutron contribution to the GDR 
state is $\approx 2$. For the GDR states the {$2qp$}
configurations predominantly correspond to excitations from the 
$sd$-shell to the $fp$-shell. The structure of the pygmy states,     
on the other hand, is dominated by transitions from the 
$1f_{7/2}$ proton state at -0.17 MeV, and from the $2p_{3/2}$ 
proton state at 3.60 MeV (this state is only bound because 
of the Coulomb barrier). The energy weighted sum of the strength 
below 11 MeV excitation energy corresponds to 2.5\% of the 
Thomas-Reiche-Kuhn sum rule (TRK).

\begin{figure}
\includegraphics[scale=0.6,angle=270]{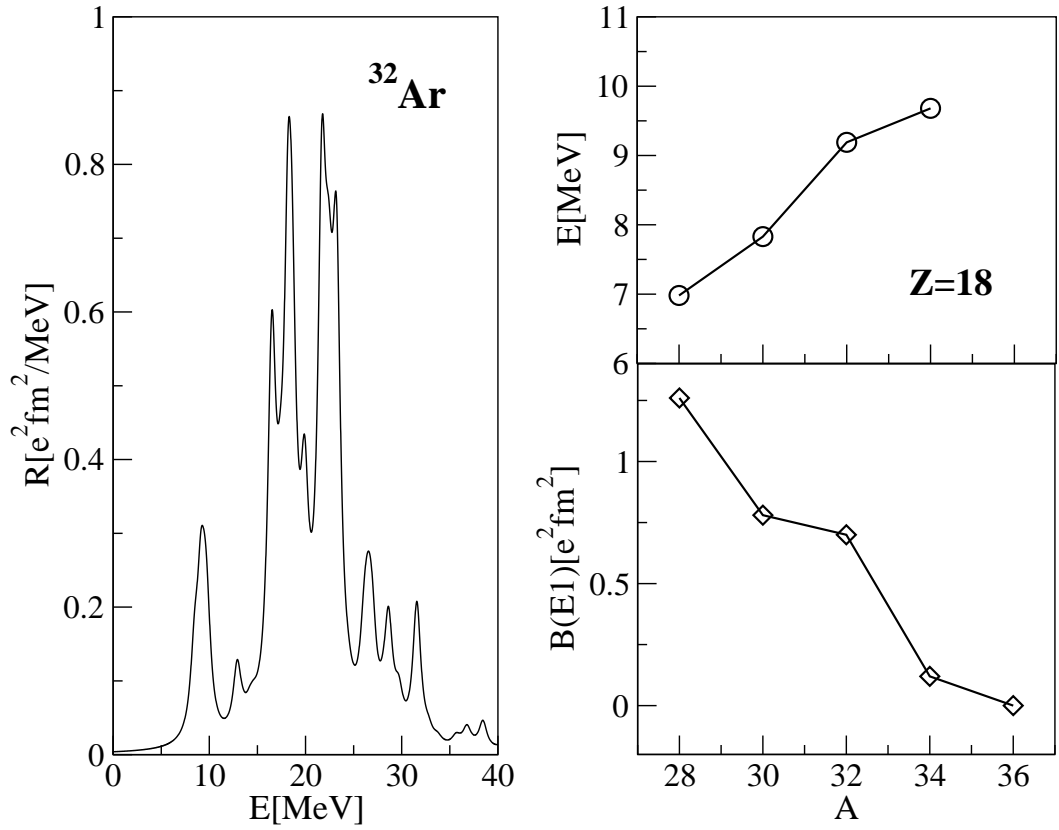}
\caption{The RHB+RQRPA isovector dipole strength distribution
in $^{32}$Ar (left panel). For the argon isotopes, 
the mass dependence of the centroid energy of the pygmy peak 
and the corresponding values of the integrated B(E1) strength 
below 10 MeV excitation energy are shown in the panel 
on the right.}
\label{fig3}
\end{figure}

\begin{figure}
\includegraphics[scale=0.75,angle=0]{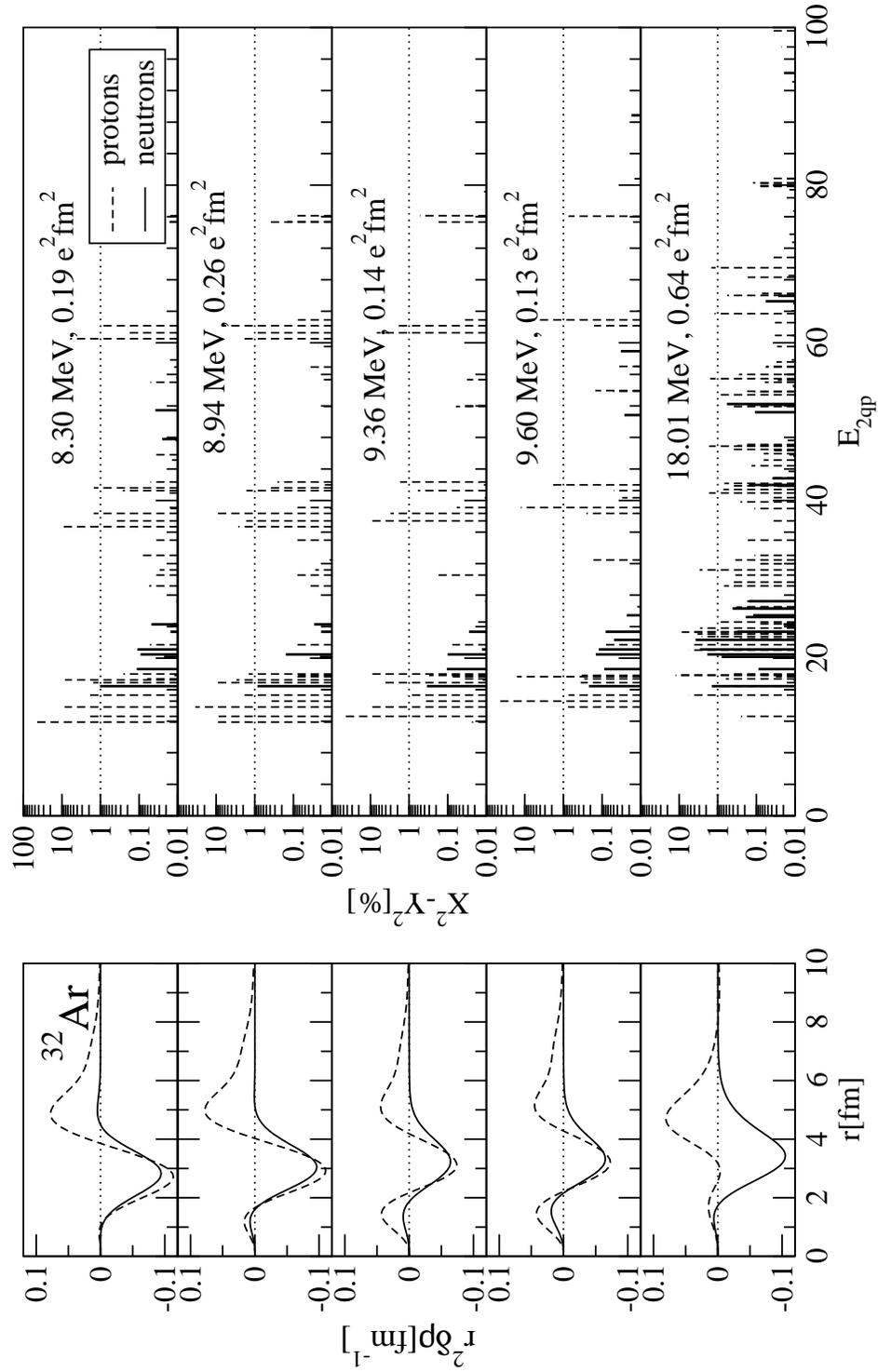}
\caption{Transition densities and RQRPA amplitudes 
for selected $1^-$ states in $^{32}$Ar.
For the main peaks in the low-energy 
region and in the region of the isovector GDR, 
the proton and neutron transition densities are 
shown in the left panel. The corresponding distributions 
of the RQRPA amplitudes are displayed as functions of the unperturbed 
energy of the respective $2qp$-configurations (right panel).
}
\label{fig4}
\end{figure}

Another example of particularly pronounced proton PDR are the 
proton-rich isotopes of Ar. In the left panel of 
Fig.~\ref{fig3} we display the RHB+RQRPA electric dipole strength 
distribution in $^{32}$Ar. In addition to the 
rather fragmented GDR structure at $\approx$ 20 MeV, prominent 
proton PDR peaks are obtained just below 10 MeV. For the four states
at 8.30, 8.94, 9.36, and 9.60 MeV, which constitute the pygmy 
structure, in Fig.~\ref{fig4} we plot the proton and neutron 
transition densities, and the distributions of the RQRPA amplitudes, 
in comparison with the GDR state at 18.01 MeV. In contrast to the 
isovector GDR, the proton and neutron 
transition densities of the pygmy states are in phase. The RQRPA amplitudes 
of the low-lying states present superpositions of many 
proton {$2qp$} configurations, with the neutron contributions 
at the level of 1\%. The dominant configurations correspond to transitions
from the proton states $1d_{3/2}$ at -2.09 MeV and $2s_{1/2}$ at -4.07 MeV.
The corresponding B(E1) values are included in Fig.~\ref{fig4}. 
The energy weighted sum of the strength below 10 MeV takes 
5.6\% of the TRK sum rule. In the right panel of Fig.~\ref{fig3}
we display the mass dependence of the centroid energy of the pygmy peaks 
and the corresponding values of the integrated B(E1) strength 
below 10 MeV excitation energy.
In contrast to the case of medium-heavy
and heavy neutron-rich isotopes, in which  
both the PDR and GDR are lowered in energy with the increase of the 
neutron number \cite{Paar.05}, 
in proton-rich isotopes the mass dependence of the PDR excitation 
energy and B(E1) strength is opposite to that of the GDR. 
The proton PDR decreases in energy with the growth of the proton excess.
This mass dependence is intuitively expected since, as we have shown,
the proton PDR mode is dominated by transitions from the weakly-bound
proton orbitals. As the proton drip-line is approached,
either by increasing the number of protons or by decreasing the number of
neutrons, due to the weaker binding of higher proton orbitals one expects
more inert oscillations, i.e. lower excitation energies. The number of
2qp configurations which include weakly-bound proton orbitals increases
towards the drip-line, resulting in an enhancement of the
low-lying B(E1) strength. 

In conclusion, we have employed the self-consistent RHB model and the 
RQRPA in the analysis of the low-energy dipole response in proton-rich
nuclei close to the proton drip-line. For nuclei with more protons than
neutrons, in the chain of N=20 isotones 
the RHB+RQRPA calculation predicts the occurrence of an exotic mode in the
low-energy region below 10 MeV: the proton electric pygmy dipole resonance.
The analysis of the corresponding proton and neutron transition densities shows
that proton PDR states correspond to the
oscillation of the proton excess against an approximately isospin-saturated core. 
The RQRPA amplitudes of the pygmy states display a superposition
of many proton {$2qp$} configurations, whereas the contribution of neutron 
excitations is at the level of 1\%. The dominant configurations
include weakly-bound proton orbitals, and also the states which are
bound only because of the presence of the Coulomb barrier.
In contrast to the rather fragmented
GDR, the PDR strength is concentrated in a narrow region of excitation
energies well below the region of giant resonances, and the
energy weighted sum of the PDR strength corresponds to a few percent 
of the TRK sum rule. With the decrease of the number of neutrons, 
for the chain of proton-rich Ar isotopes the proton PDR is lowered 
in energy, and the integrated B(E1) strength
in the low-energy region increases accordingly. 
For heavier nuclei the proton drip-line is located in the region of 
neutron-deficient, rather than proton-rich nuclei, and therefore 
no low-lying dipole strength is calculated in medium-heavy and 
heavy nuclei close to the proton drip-line. 

\bigskip 
\leftline{\bf ACKNOWLEDGMENTS}

This work has been supported in part by the Bundesministerium
f\"ur Bildung und Forschung - project 06 MT 193, 
by the Deutsche Forschungsgemeinschaft (DFG) under contract SFB 634,
by the Croatian Ministry of Science - project 0119250, 
and by the Gesellschaft f\" ur Schwerionenforschung (GSI) Darmstadt.

\bigskip \bigskip

\newpage

\end{document}